\documentclass{aa}
\usepackage{color}
\usepackage{graphicx}
\usepackage{amsmath}
\usepackage{amssymb}
\usepackage[T1]{fontenc}
\usepackage{mathptmx}
\usepackage{ae,aecompl} 
\usepackage[varg]{txfonts}

\begin{document}
\title{Reconciling inverse-Compton Doppler factors with variability Doppler factors in blazar jets}
\author{I. Liodakis\inst{\ref{inst1},}\inst{\ref{inst2}}\thanks{liodakis@physics.uoc.gr}
\and A. Zezas\inst{\ref{inst1},}\inst{\ref{inst2}}
\and E. Angelakis \inst{\ref{inst3}}
\and T. Hovatta \inst{\ref{inst4},}\inst{\ref{inst5},}\inst{\ref{inst6}}
\and V. Pavlidou\inst{\ref{inst1},}\inst{\ref{inst2}}
}

\institute{Department of Physics and ITCP \thanks {Institute for Theoretical
and Computational Physics, formerly Institute for Plasma Physics}, University of Crete, 71003, Heraklion, Greece\label{inst1}
\and Foundation for Research and Technology - Hellas, IESL, Voutes, 7110 Heraklion, Greece \label{inst2}
\and Max-Planck-Institut f\"ur Radioastronomie, Auf dem H\"ugel 69, 53121 Bonn, Germany \label{inst3}
\and Aalto University Mets\"ahovi Radio Observatory, Mets\"ahovintie 114, 02540 Kylm\"al\"a, Finland\label{inst4}
\and Aalto University Department of Radio Science and Engineering,P.O. BOX 13000, FI-00076 AALTO, Finland \label{inst5}
\and Tuorla Observatory, Department of Physics and Astronomy, University of Turku, Finland\label{inst6}
}

\abstract{Blazar population models have shown that the inverse-Compton and variability Doppler factor estimates yield consistent results at the population level for flat spectrum radio quasars (FSRQs). The two methods, however, are inconsistent when compared on a source-by-source basis.}{In this work, we attempt to understand the source of the discrepancy by tracing the potential sources of systematic and statistical error for the inverse-Compton Doppler factors. By eliminating these sources of error, we provide stronger constrains on the value of the Doppler factor in blazar jets. } {We re-estimate the inverse-Compton Doppler factor for 11 sources that meet certain criteria for their synchrotron peak frequency and the availability of Doppler factor estimates in the literature. We compare these estimates with {the average of two} different estimates of the variability Doppler factor obtained using various datasets and methodologies\ to identify any discrepancies and, in each case, trace their sources in the methodology or assumptions adopted.}{We identify three significant sources of error for the inverse-Compton Doppler factors: a) contamination of the X-ray flux by non-synchrotron self-Compton emission; b) radio observations at frequencies other than the synchrotron turnover frequency; c) non-simultaneity between radio and X-ray observations. We discuss key aspects in the correct application of the inverse-Compton method in light of these potential errors. We are able to constrain the Doppler factor of {3C273, 3C345, 3C454.3, PKS1510-089, and PKS1633+382} effectively, since all available estimates from both methods converge to the same values for these five sources.}{}


\titlerunning{Reconciling $\delta_{IC}$ with $\delta_{var}$ in blazar jets.}
\authorrunning{Liodakis et al.}
\maketitle

\section{Introduction}\label{introd}

Blazars, and in particular  flat spectrum radio quasars (FSRQs) and BL Lac objects, are active galactic nuclei (AGN) with jets closely aligned to our line of sight \citep{Readhead1978,Blandford1979,Scheuer1979,Readhead1980}. They constitute one of most interesting classes of AGN owing to the relativistic effects dominating their broadband emission, thereby complicating our understanding of their intrinsic properties. Revealing the properties of blazars and their jets in their rest frame would allow us to study important astrophysical processes in supermassive black hole jets, including emission mechanisms, and jet production, collimation, propagation, and energetics.

A major difficulty in the study of blazar jets in their rest frame is the limitation on obtaining reliable estimates of the Doppler factor in blazar jets, i.e., of the amount of relativistic boosting. Doppler factors are notoriously hard to estimate and the lack of confident estimates is known to hinder the identification of potentially revealing empirical correlations between rest-frame blazar properties (e.g., \citealp{Hovatta2010,Lister2011,Blinov2016-II,Blinov2016}, and \citealp{Angelakis2016}).

Several methods have been proposed for the estimation of Doppler factors. Very often different methods rely on different assumptions regarding the physical properties of the jet, and in many cases these methods produce different results. The methods commonly used include the inverse-Compton method (\citealp{Ghisellini1993}, hereafter G93), which assumes that synchrotron self-Compton (SSC) is the dominant emission mechanism at X-ray frequencies, and methods relying on the assumption of equipartition between radiating particles and magnetic field \citep{Readhead1994}. These latter methods include the equipartition Doppler factor \citep{Readhead1994,Guijosa1996} and the variability Doppler factor methodologies \citep{Valtaoja1999,Lahteenmaki1999-III,Hovatta2009,Liodakis2017}. {The different assumptions used by these methods make a direct comparison unfeasible. The situation is further complicated if the Doppler factor does not remain constant in time. Local acceleration, jet precession, and bents in the jet could in principle yield different results for the same source depending on the time of the observations.

\cite{Liodakis2015-II} have been able to evaluate these methods in a statistical fashion via population models they optimized in \cite{Liodakis2015}.} These models consist of distributions for the Lorentz factor and the intrinsic monochromatic luminosity for different blazar classes optimized to reproduce the observed apparent velocity and redshift distributions of the MOJAVE (Monitoring of Jets in Active galactic nuclei with VLBA Experiment; \citealp{Lister2005})\footnote{http://www.physics.purdue.edu/MOJAVE/} sample. Using these population models, the Doppler factor distributions of the FSRQs and BL Lac objects have been produced through Monte Carlo simulations {assuming the continuous jet case}. 

\cite{Liodakis2015-II} have compared these model distributions with the distributions of Doppler factors estimated through each of the methods above available in the literature, accounting for sample size and flux limit. These authors have found that both the inverse-Compton ($\delta_{IC}$) and the variability ($\delta_{var}$) Doppler factor methods can adequately describe the FSRQs population. For the inverse-Compton (IC) Doppler factors, an error analysis at the population level has shown that their errors are normally distributed and that each estimate has a  $\sim 63\%$ error on average. A Kolmogorov-Smirnov test (K-S test) gave a $\sim 95\%$ probability of consistency between the observed and simulated (with errors) distributions. In addition, not only is the statistical error of the IC Doppler factors quite high, but there are also many potential sources of systematic error in the estimates as a result of the assumptions involved in the method. Individual estimates for the variability Doppler factors (\citealp{Hovatta2009}; hereafter H09)
were found to have on average a $\sim 30\%$ statistical error, making them more accurate, however the dominant source of error in this case is systematic and due to the finite cadence of observations. Sources most likely to be affected by this type of systematic uncertainty can be identified by comparing the cadence of observations to the fastest flare detected. In the remaining sources, the dominant source of error is the $30\%$ {statistical} error that is due to the uncertainty in estimating the rise time and amplitude of a flare.

Recently, a different approach allowed for a more accurate estimation of $\delta_{var}$. By modeling the multiwavelength radio light curves from the F-GAMMA program\footnote{http://www3.mpifr-bonn.mpg.de/div/vlbi/fgamma/fgamma.html} \citep{Fuhrmann2007,Angelakis2010} and an upgraded version of the algorithms introduced in \cite{Angelakis2015}, it was possible to estimate the $\delta_{var}$ for 58 sources with  an on-average 16\% error  (\citealp{Liodakis2017}; hereafter L17). This approach can, in addition, mitigate the effects of limited cadence and provide error estimates on a source-by-source basis.

Since both methods (IC and variability Doppler factors) can adequately describe the FSRQ population, a comparison on a blazar-by-blazar basis between the two should, in principle, allows us to control the (various) systematics that affect each method and effectively constrain the  Doppler factor value in blazar jets for which both estimates are available. However, such a comparison of the two methods yields inconsistent results even when accounting for the large statistical errors of IC Doppler factors.

In this work, we attempt to trace the sources of these discrepancies and reconcile Doppler factor estimates produced by these two approaches. In the case of 
variability Doppler factors, the availability of two independent estimates facilitates systematics control and the bracketing of possible values. In the case of IC Doppler factors,
we investigate possible sources of error and produce updated estimates by reducing, as far as possible, these sources of uncertainty.

This paper is organized as follows: In section \ref{Doppl_fact_Estim} we provide a brief description of the inverse-Compton and variability methods for Doppler factor estimation. In section \ref{sources_of_error} we analyze the potential sources of error in estimating the $\delta_{IC}$. In section \ref{error_analysis} we discuss the selection criteria for the sources for which we compare the different Doppler factor estimates and describe our methodology for producing new estimates of $\delta_{IC}$. In section \ref{res+disc} we discuss our results and their implications of obtaining improved Doppler factor estimates of relativistic blazar jets, and in section \ref{summary} we summarize our findings.

Throughout this work we have adopted $H_0=71$ ${\rm km \, s^{-1} \, Mpc^{-1}}$, $\Omega_m=0.27$ and $\Omega_\Lambda=1-\Omega_m$ \citep{Komatsu2009}.

\section{Methods for Doppler factor estimation}\label{Doppl_fact_Estim}
A lower limit on the Doppler factor can be derived by acknowledging that the SSC flux density cannot  exceed the observed flux density at high frequencies. The SSC emission is produced by the IC upscattering of synchrotron photons by the same relativistic electrons that produced these photons. The inverse-Compton Doppler factor $\delta_{IC}$ \citep{Ghisellini1993} is derived on the assumption that {\em all}
of the observed flux density at X-ray frequencies is, in fact, due to the SSC process.

Assuming a homogeneous magnetic field, and that the energy distribution of the electrons follows a power law, the Doppler factor would be
\begin{equation}
\delta_{IC}=f(\alpha)F_m\left[\frac{ln(\nu_b/\nu_m)}{F_\chi\theta_d^{6+4\alpha}\nu_\chi^\alpha\nu_m^{5+3a}}\right]^{1/(4+2\alpha)}(1+z),
\label{delta_ic}
 \end{equation} 
where $F_m$ is the synchrotron flux density of the core at frequency $\nu_m$ and  $F_\chi$ is the X-ray flux density, both in Jy; $\theta_\mathrm{d}$ is the angular size of the core in milliarcsec, $\nu_\chi$ is the X-ray observations energy in keV,  $\nu_m$ is the radio observations frequency in GHz; and $\nu_b$ is the synchrotron high-energy cutof,f which is assumed to be $10^{14}$ Hz. The function $f(\alpha)$ is given by $f(\alpha)\simeq 0.08\alpha+0.14,$ where $\alpha$ is the optically thin spectral index \citep{Ghisellini1987} assumed to be $\alpha=0.75$ . Equation \ref{delta_ic} is for the discrete jet case. For the continuous case the $\delta_{IC}$ is transformed as
\begin{equation}
\delta_{cont}=\delta_{discr}^{(4+2\alpha)/(3+2\alpha)}.
 \end{equation} 
A detailed description of the method can be found in \citet{Ghisellini1993} and \citet{Guijosa1996}.

The variability Doppler factor ($\delta_{var}$) uses the evolution of a flare in the time domain to calculate the brightness temperature of the emission region. Either by fitting exponential curves (H09) or multiwavelength modeling (L17), one can calculate the observed variability brightness temperature. Assuming that during a flare the intrinsic brightness temperature is equal to the equipartition brightness temperature \citep{Readhead1994}, the Doppler factor is proportional to the cube root of the ratio of the variability and equipartition brightness temperatures (Eq. \ref{delta_var}),
i.e., \begin{equation}
\delta_{var}\propto\left(\frac{T_{b,var}}{T_{eq}}\right)^{1/3}
\label{delta_var}
 .\end{equation} 
For a detailed description of both approaches see \citet{Valtaoja1999,Lahteenmaki1999-II,Lahteenmaki1999-III,Hovatta2009,Angelakis2015}, and L17.

\section{Sources of error for the inverse-Compton Doppler factors}\label{sources_of_error}

Owing to the physics and the assumptions involved in the method, there are three main sources of error in inverse-Compton Doppler factor estimates: the presence of external X-ray emission (not due to SSC), radio observations performed at a frequency that is different than the spectrum turnover frequency, and the lack of simultaneity between X-ray and radio observations.

\subsection{External X-ray flux} 

As pointed out above and discussed by \cite{Britzen2007},{ the $\delta_{IC}$ method } can produce a reliable estimate only if all of the measured X-ray emission is caused by inverse-Compton scattering of synchrotron photons. Potential sources of the non-SSC X-ray flux could be the hot corona of the accretion disk, external Compton scattering of the broad-line region photon field, or synchrotron radiation of intermediate or high-peaked synchrotron sources.  For this reason, as shown in \cite{Liodakis2015-II}, the method fails to adequately describe the BL Lac population, for which the synchrotron peak reaches keV energies contaminating the X-ray flux. In addition, orphan X-ray flares (without a radio counterpart) might also indicate external origin of the X-ray flux. Sources with a prominent big blue bump also have to be treated with  caution because of the output of the bump that possibly, but not necessarily, extends to the X-ray regime. Any one of these effects, or their combination, could lead to underestimating $\delta_{IC}$ and introducing systematic errors. Thus in the application of the method, careful sample selection is of paramount importance.

\subsection{Observed versus turnover radio frequency}\label{obs_turn}

In G93 the authors use the observed flux density and frequency for the estimation of the $\delta_{IC}$. It is argued in \cite{Lahteenmaki1999-II} that this is an approximation in the application and that the method requires the turnover frequency (i.e., the frequency at which the synchrotron spectrum changes from optically thick to optically thin) and flux density to be used instead. The counter-argument for this criticism is that at any given frequency, the VLBI observations show self-absorbed regions. Investigation of this effect is not trivial. Measurements at the turnover frequency are only available for a handful of sources. In addition, although the turnover is usually located in high frequencies (>43 GHz), this is not always the case \citep{Rabaca1994,Fromm2015} and is different for every blazar. The location of the turnover frequency has also been shown to change, moving either to higher or lower frequencies during outbursts \citep{Feng2006,Fromm2015}. Recent results from the F-GAMMA survey show that the turnover frequency can range from $<2.6$ to $>86$ GHz in a single source (Angelakis et al. in prep., also \citealp{Angelakis2012}). In order to estimate the impact of this approximation on Doppler factor estimation, data at multiple radio frequencies are necessary. Such data are available for a few sources we study in this work, and they indicate that this effect can play a crucial role in the correct application of the method.

\subsection{Simultaneity of observations} 

The radio and X-ray observations should be
contemporaneous since the SSC model requires the relativistic electrons producing the radio flux to up-scatter the synchrotron photons they themselves create to higher energies that produce the X-ray flux. Owing to the variable nature of blazars, a significant time difference between radio and X-ray observations can result in different initial emission conditions and thus can lead to either underestimation or overestimation of the ``true'' Doppler factor. Past estimates of the inverse-Compton Doppler factor have not systematically pursued the use of simultaneous X-ray and radio observations, making the lack of simultaneity an obvious candidate for a source of error in the estimates. Since this effect can affect estimates in either direction, it will contribute to the statistical error budget. The G93 sample was found to be dominated by statistical error (63\% on average on each estimate). How much of that 63\% error can the lack of simultaneity between X-ray and radio observations account for is still to be determined, and the new $\delta_{IC}$ estimates produced in this work using contemporaneous radio and X-ray measurements are a major step in this direction.

\section{Sample and data}\label{error_analysis}
\begin{figure}
\resizebox{\hsize}{!}{\includegraphics[scale=1]{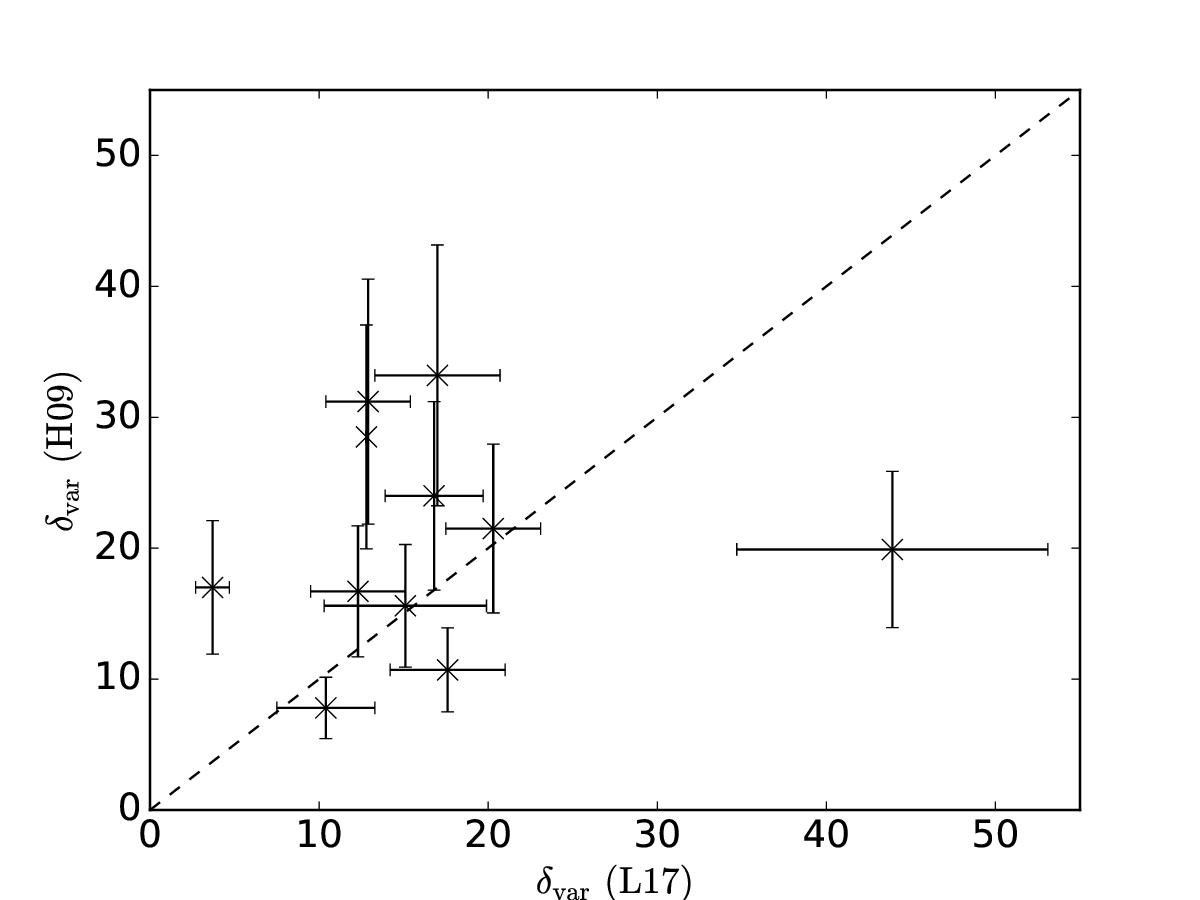} }
\caption{Comparison between the $\delta_{var}$ from H09 with the $\delta_{var}$ from L17. The dashed line represents equality of the estimates. The error bars on the vertical axis are the on-average 30\%  from {\cite{Liodakis2015-II}.}}
\label{plt:H09_vs_L16}
\end{figure}  

The sources of error discussed above, which were all entered through violations of basic assumptions underlying the inverse-Compton Doppler factor estimation method, are expected to dominate over observational errors in the quantities entered in Eq. \ref{delta_ic}. We adopted the following approach to form a quantitative understanding of the uncertainties in $\delta_{IC}$ and resolve the discrepancy with variability Doppler factors:

\begin{itemize}
\item {We selected a sample of sources to study for which two different $\delta_{var}$ estimates, with various methodologies, are available (from both L17 and H09) so that we could control whether the $\delta_{var}$ , for which we are comparing $\delta_{IC}$ estimates against, are likely to be considerably plagued by systematics. Figure \ref{plt:H09_vs_L16} shows the comparison between the $\delta_{var}$ estimates. Differences between the variability estimates do not necessarily suggest problems with the $\delta_{var}$. Differences could arise from the different time span of observations between the datasets used especially if the Doppler factor varies with time. Exceptional outbursts or fast flares may occur in one dataset but (equivalent events) not in the other, leading to one producing a larger value for the $\delta_{var}$. Discrepancies between the variability estimates are discussed in more detail in L17. }
\item We focused on FSRQs (low-synchrotron-peaked sources) to minimize contamination of the X-ray flux by non-SSC sources by at least eliminating the possibility of synchrotron emission at X-ray frequencies. 
\item We sought  contemporaneous X-ray and radio data in the literature and used them to recalculate $\delta_{IC}$, thus eliminating the error induced by the lack of simultaneity between X-ray and radio observations.
\item We {re-examined} any residual discrepancy between recalculated $\delta_{IC}$ and $\delta_{var}$, and we evaluated this discrepancy on a source-by-source basis. 
\end{itemize}

We began our source selection from the F-GAMMA sample, for which new estimates of $\delta_{var}$ have been obtained by L17\footnote{Our selection of F-GAMMA sources also results in a sample of high-interest blazars, which, for example, are $\gamma$-ray loud \citep{Acero2015} and are regularly monitored in optical polarization \citep{King2014,Pavlidou2014}.}. F-GAMMA is a multiwavelength monitoring program of the most interesting $\gamma$-loud sources with high radio power detected by the Large Area Telescope (LAT) on board the Fermi Gamma-ray Space Telescope \citep{Acero2015}. 

We limited ourselves to a subsample of F-GAMMA sources that (a) are low-synchrotron-peaked FSRQs \citep{Giommi2012-II}, (b) have a previous estimation of both the $\delta_{IC}$ and $\delta_{var}$ (H09) available, (c) have available quasi-simultaneous (less than a week apart) VLBI and X-ray observations in the literature. There are 11 sources that meet these criteria. These sources are PKS0420-014, PKS0528+134, PKS1156+295, PKS1510-089, PKS1633+382, PKS1730-130, 3C273, 3C279, 3C345, 3C454.3, and CTA102.

The X-ray observations for PKS0528+134 were performed by the XMM-Newton observatory \citep{Jansen2001}, while the radio observations were performed as part of the Boston University group (BU group)\footnote{http://www.bu.edu/blazars} monthly monitoring program \citep{Palma2011}. The X-ray observations for 3C273 were performed by the ROSAT X-ray observatory, whereas the radio observations by the Very Long Baseline Interferometry (VLBI) were performed in a multifrequency campaign \citep{Mantovani2000}. The X-ray observations for PKS1510-089 and PKS1633+382 were conducted by the Rossi X-Ray Timing Explorer (RXTE) and the radio observations were conducted by the VLBA (\cite{Marscher2010}, and \cite{Jorstad2011}, respectively). 

The X-ray flux for the remaining sources was taken from \cite{Chang2010} (Swift X-ray telescope), while we used either \cite{Lister2013} or the data available online as part of the monitoring program by the BU group for the radio observations. For the data from the monitoring program by the BU group, we used standard DIFMAP procedures and the clean model available for each source for the phase and amplitude calibration. We then proceeded to fit Gaussian components to the calibrated maps. For our calculations, we use only the flux density of the core.

{Determining the state of each source at the time of the observations is not trivial. A source might appear quiescent in one frequency (e.g., radio) while undergoing outbursts in another (e.g., optical). Unfortunately, the multiwavelength information necessary to characterize the state of a source is not always available. Regarding the sources for which we have that information, PKS0528+134 is in quiescent, PKS1510-089 is in outburst in multiple  wavebands, and PKS1633+382 appears to be in quiescent in radio and X-rays while in outburst in $\gamma$-rays.}

We converted  all the broadband fluxes available in the literature to monochromatic flux densities at 1 keV required for Eq. \ref{delta_ic}, using the best-fit power-law spectrum for each source, and assuming the Galactic line-of-sight atomic hydrogen column density $(N_H)$ calculated from the survey of \cite{Dickey1990} via the Colden CIAO tool. In the cases for which we lacked spectral information, the X-ray photon index was taken from \cite{Williamson2014}. All our Doppler factor estimates were calculated for the continuous jet case. The reason for this is that in \cite{Liodakis2015-II} we found that the continuous jet case provided better agreement with blazar data on a population level.
\begin{table*}
\setlength{\tabcolsep}{11pt}
\centering
  \caption{ {\bf Data.} Column: (1) Source name, (2) Redshift (3) Radio frequency, (4) X-ray flux density at 1 keV ($F_{1keV}$), (5) radio flux density ($F_R$), and (6) angular size ($\theta_d$) of the core used in the estimation of the Doppler factor for each source.}
  \label{tab:flux_values}
\begin{tabular}{@{}lccccc@{}}
 \hline
   Source & Redshift & Radio frequency &$F_{1keV}$ &$F_{R}$ & $\theta_d$ \\ 
   & & (GHz) & ({$\mu$}Jy)  &  (Jy) & (mas) \\
   \hline 

  PKS0420-014 & 0.915 &15 & 0.61 & 4.26 & 0.306 \\ 
  PKS0528+134 & 2.070 & 43 & 0.06 & 0.84 & 0.048  \\ 
  PKS1156+295 & 0.729 & 43 & 0.29 & 1.49 & 0.1 \\ 
  PKS1510-089 & 0.360 & 15 & 1.44  & 3.80 & 0.2   \\ 
  PKS1633+382 & 1.814 & 43 & 0.30 & 1.60 & 0.07 \\ 
  PKS1730-130 & 0.902 &15 & 0.20 & 2.54 & 0.144  \\ 
  PKS1730-130 & 0.902 &43 & 0.20 & 0.9 & 0.126 \\ 
  3C273 & 0.158 & 43 & 4.50${\times}10^{-7}$ & 2.57 & 0.42  \\ 
  3C279 & 0.536 & 15 & 1.18 & 13.3 & 0.18  \\ 
  3C345 & 0.593 & 15 & 7.40${\times}10^{-3}$  & 5.09 & 0.165 \\
  3C345 & 0.593 & 43 & 7.40${\times}10^{-3}$ & 3.55 & 0.221 \\ 
  3C454.3 & 0.859 & 15 & 3.18 & 22.20 &  0.45 \\
  3C454.3 & 0.859 & 43 & 3.18 & 20.30 & 0.09   \\ 
  CTA102 & 1.037 & 43 & 0.39 & 2.40 & 0.2 \\ 
\hline
\multicolumn{5}{c}{(\citealp{Mantovani2000,Jansen2001,Marscher2010,Jorstad2011};}\\
\multicolumn{5}{c}{\citealp{Palma2011,Chang2010,Lister2013})}\\
\end{tabular}
\end{table*}  

All the values required for the estimation of the $\delta_{IC}$ are summarized in Table \ref{tab:flux_values}. {For the synchrotron high-energy cutoff and the optically thin spectral index we assumed $\nu_b=10^{14}$ Hz and $\alpha=0.75,$ respectively, to match the values used in G93. We verified that the method is not significantly affected by the choice of these two values. Any $\nu_b$ in the range of $10^{11}-10^{17}$ Hz results in at most a $\sim 10\%$ fractional difference from the estimated value (Table \ref{tab:doppler_est}), while any $\alpha$ in the range of $0.55-0.95$ results in at most a $\sim 20\%$ fractional difference.
}

\section{Results}\label{res+disc}
\begin{figure}
\resizebox{\hsize}{!}{\includegraphics[scale=1]{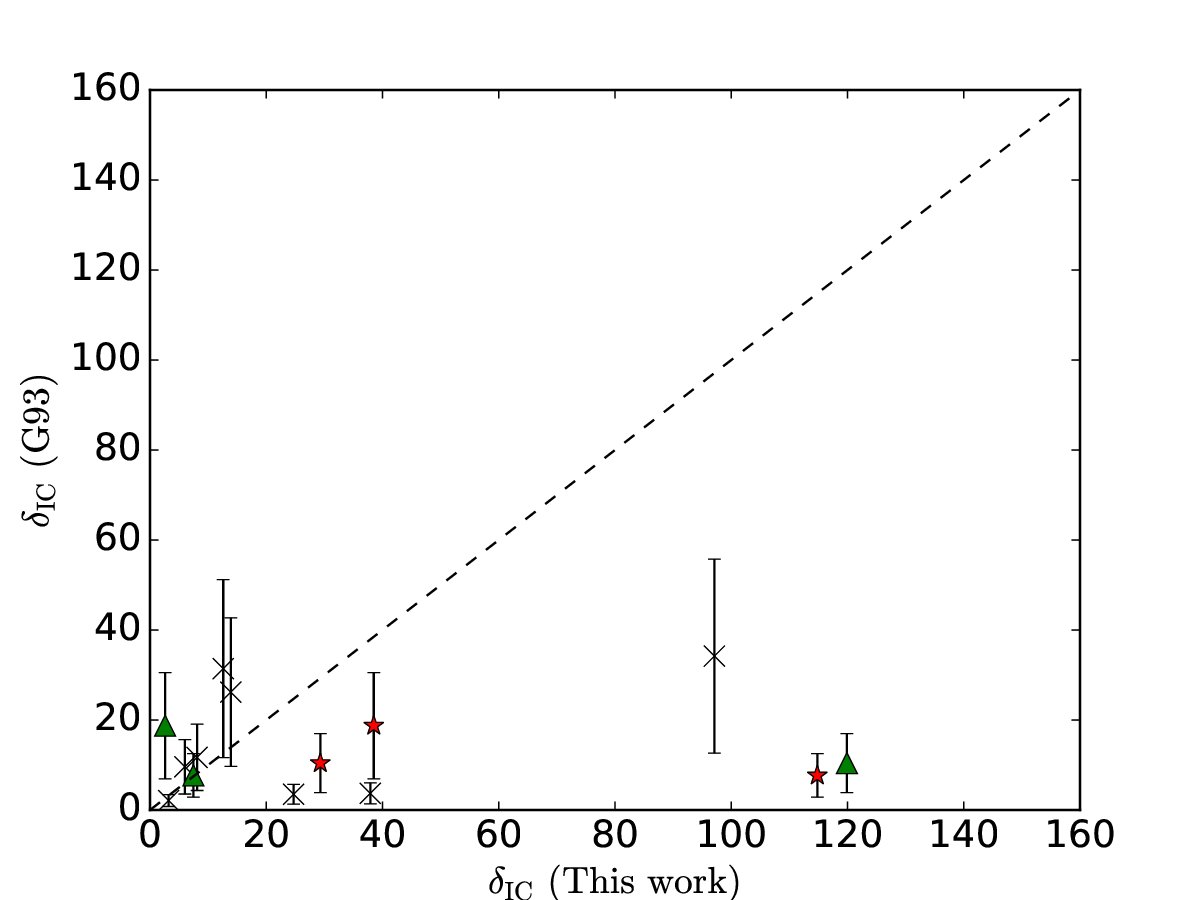} }
\caption{Comparison between the $\delta_{IC}$ from this work with the $\delta_{IC}$ from G93. The red $\star$ indicates the 15 GHz, whereas the green triangles indicate the 43 GHz values for the three sources with  two $\delta_{IC}$ estimates. The dashed line represents the equality of the estimates. The error bars on the vertical axis show the on-average 63\% {from \cite{Liodakis2015-II}.}}
\label{plt:thiswork_vs_G93}
\end{figure}

\begin{figure}
\resizebox{\hsize}{!}{\includegraphics[scale=1]{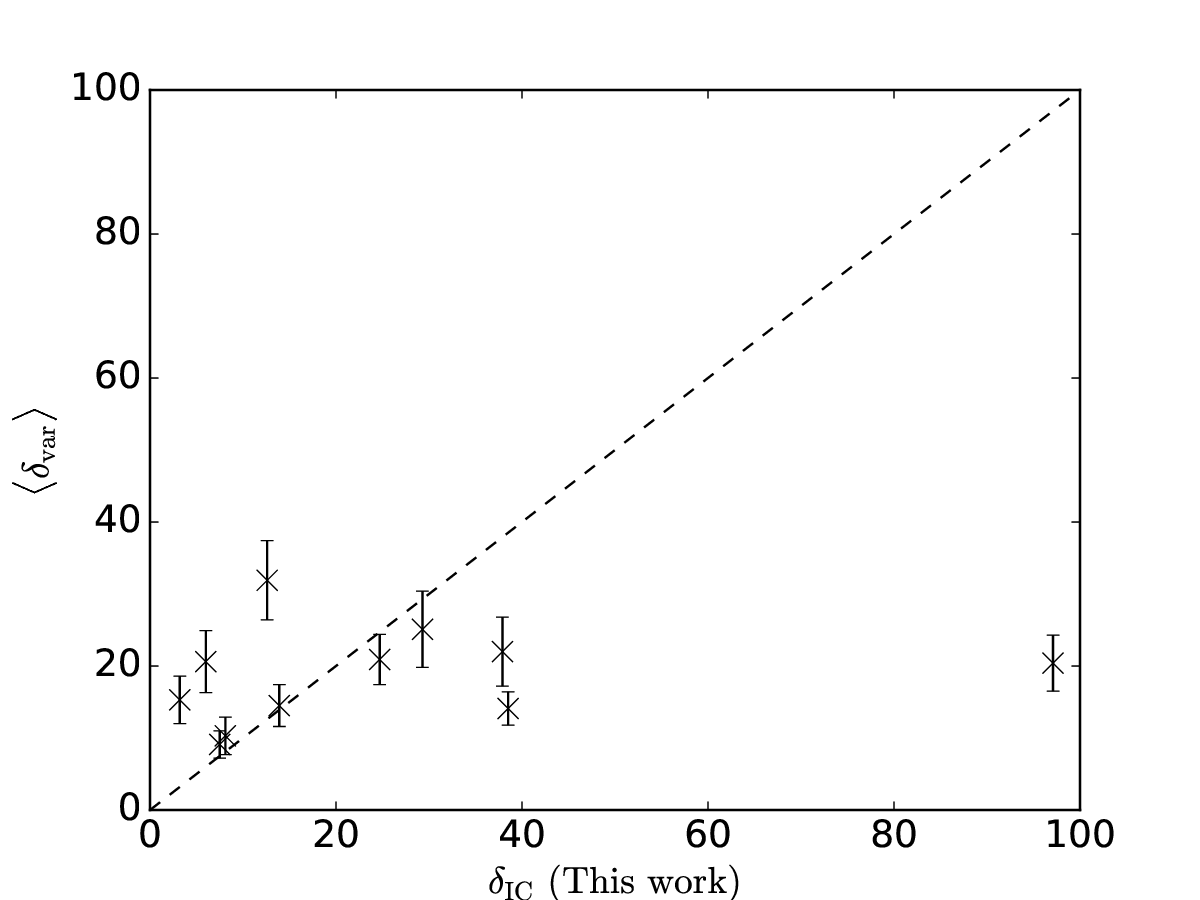} }
\caption{Comparison between the $\delta_{IC}$ from this work with the mean of the two estimates in the literature ($\langle\delta_{var}\rangle$, H09, L17). The dashed line represents equality of the estimates and the errors in the vertical axis represent the spread between the two methods.  For the sources with estimates for two frequencies, we plot the estimate closest to $\langle\delta_{var}\rangle$.}
\label{plt:thiswork_vs_H09-L16}
\end{figure}

Using the values from Table \ref{tab:flux_values} and Eq. \ref{delta_ic} we re-estimated $\delta_{IC}$ for the 11 sources in our sample. Radio observations at both 15 and 43 GHz were available for sources
3C345, 3C454.3, and PKS1730-130.

{Figures \ref{plt:thiswork_vs_G93} and \ref{plt:thiswork_vs_H09-L16} show} the comparison between the $\delta_{IC}$ derived in this work with various estimates in the literature. For the variability Doppler factors (Fig. \ref{plt:thiswork_vs_H09-L16}) we plot the mean $\pm 0.5\times$ (difference between H09, L17), since it quantifies our systematic uncertainty in the $\delta_{var}$ estimate. In figure \ref{plt:thiswork_vs_H09-L16} for the sources with estimates for two radio frequencies, we only plot the estimate closest to the variability estimates. The various Doppler factor estimates are summarized in Table \ref{tab:doppler_est}.

\begin{table*}
\setlength{\tabcolsep}{11pt}
\centering
  \caption{{\bf Doppler factor estimates.} Column: (1) Source name, (2) Frequency, (3) $\delta_{IC}$  derived in this work, (4) $\delta_{IC}$  from G93, (5) $\delta_{var}$ from H09, (6) $\delta_{var}$ from L17, (7) average value of the $\delta_{var}$ estimates, and (8) the spread of the $\delta_{var}$ estimates. The errors in column 4 and 5 are the 63\% and 30\% errors, respectively, {from \cite{Liodakis2015-II}, the errors in column 6 are the source-by-source estimates from L17, and the errors in column 7 are derived through standard error propagation.}}
  \label{tab:doppler_est}
\begin{minipage}{180mm}
\begin{tabular}{@{}lcccccccc@{}}
 \hline
   Source & Frequency & $\delta_{IC}$ & $\delta_{IC}$ & $\delta_{var}$ & $\delta_{var}$ &  $\langle \delta_{var} \rangle $ & $\sigma_{\langle \delta_{var} \rangle}$\\ 
           & (GHz) & (This work)  &  (G93) & (H09) & (L17)& &\\
  \hline   
PKS0420-014 & 15 & 12.6 & 31.4 $\pm$ 19.8 & 19.9 $\pm$ 5.9 & 43.9 $\pm$ 9.2 & 31.9 $\pm$ 5.5 & 12.0\\ 
PKS0528+134 & 43 & 37.9 & 3.7 $\pm$ 2.3 & 31.2 $\pm$ 9.3 & 12.9 $\pm$ 2.5 & 22.0 $\pm$ 4.8 & 9.2\\ 
PKS1156+295 & 43 & 6.0 & 9.6 $\pm$ 6.0 & 28.5 $\pm$ 8.5 & 12.8 $\pm$ 0.04 & 20.6$\pm$ 4.3 & 7.9 \\
PKS1510-089 & 15 & 13.9 & 26.2 $\pm$ 16.5 & 16.7 $\pm$ 5.0 & 12.3 $\pm$ 2.8 & 14.5$\pm$ 2.9 & 2.2 \\ 
PKS1633+382 & 43 & 24.7 & 3.5 $\pm$ 2.2 & 21.5 $\pm$ 6.4 & 20.3 $\pm$ 2.8 & 20.9$\pm$ 3.5 & 0.6 \\ 
PKS1730-130 & 15 & 38.5 & 18.7 $\pm$ 11.8 & 10.7 $\pm$ 3.2 & 17.6 $\pm$ 3.4 & 14.1$\pm$2.3 & 3.5\\ 
PKS1730-130 & 43 & 2.6 & 18.7 $\pm$ 11.8 & 10.7 $\pm$ 3.2 & 17.6 $\pm$ 3.4 &  14.1$\pm$2.3 & 3.5\\  
3C273 & 43 & 8.1 & 11.7 $\pm$ 7.4 & 17.0 $\pm$ 5.1 & 3.7 $\pm$ 1.0 & 10.3$\pm$ 2.6 & 6.7\\ 
3C279 & 15 & 97.1 & 34.2 $\pm$ 21.5 & 24.0 $\pm$ 7.2 & 16.8 $\pm$ 2.9  & 20.4$\pm$ 3.9 & 3.6 \\ 
3C345 & 15 & 114.8 & 7.6 $\pm$ 4.8 & 7.8 $\pm$ 2.34 & 10.4 $\pm$ 2.9 & 9.1$\pm$1.9 & 1.3\\ 
3C345 & 43 & 7.5 & 7.6 $\pm$ 4.8 & 7.8 $\pm$ 2.34 & 10.4 $\pm$ 2.9 & 9.1$\pm$1.9 & 1.3\\ 
3C454.3 & 15 & 29.3 & 10.4 $\pm$ 6.5 & 33.2 $\pm$ 9.9 & 17.0 $\pm$ 3.7 & 25.1$\pm$5.3& 8.1\\ 
3C454.3 & 43 & 119.9 & 10.4 $\pm$ 6.5 & 33.2 $\pm$ 9.9 & 17.0 $\pm$ 3.7 & 25.1$\pm$5.3& 8.1\\ 
CTA102 & 43 & 3.2 & 2.1 $\pm$ 1.3 & 15.6 $\pm$ 4.6 & 15.1 $\pm$ 4.8 & 15.3$\pm$3.3 & 0.25\\ 
\hline
\end{tabular}
\end{minipage}
\end{table*}  

\begin{figure}
\resizebox{\hsize}{!}{\includegraphics[scale=1]{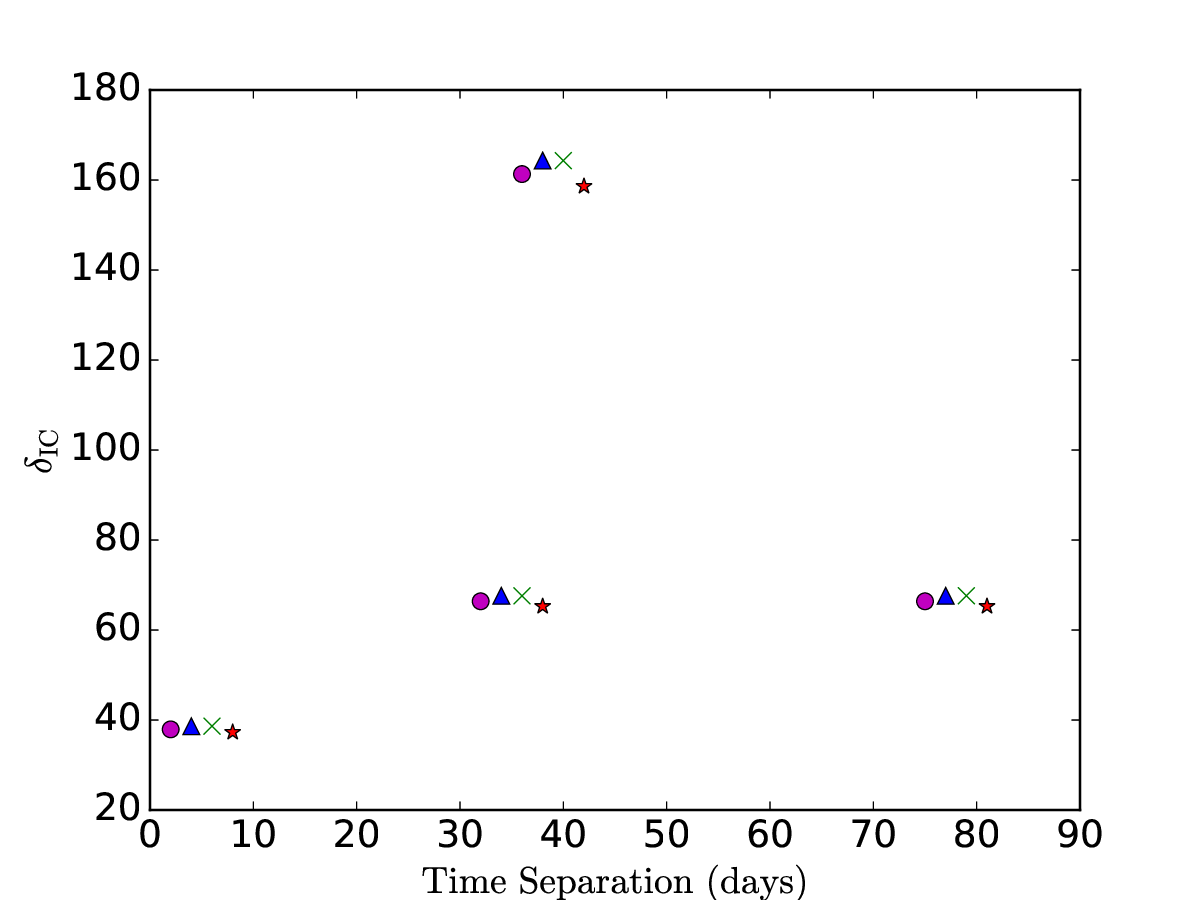} }
\caption{Various estimates for PKS0528+134 against the time separation of X-ray and radio observations. Same symbols denote same X-ray but different radio observations.}
\label{plt:frac_dif}
\end{figure} 
\begin{figure}
\resizebox{\hsize}{!}{\includegraphics[scale=1]{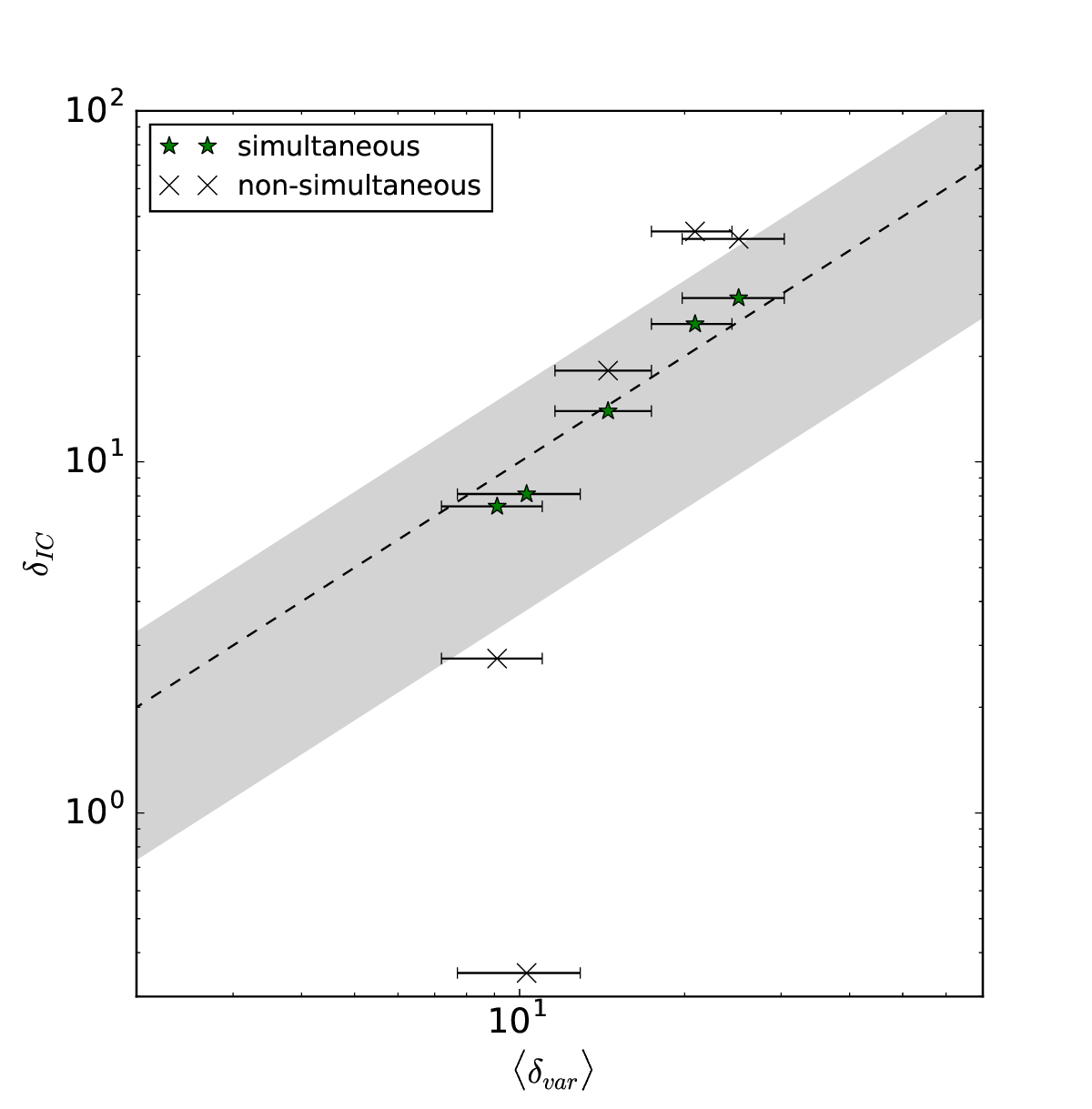} }
\caption{Comparison of the $\langle \delta_{var} \rangle$ against simultaneous and non-simultaneous estimates using the X-ray flux density {from this work and from G93, respectively.} The green $\star$ indicates for simultaneous estimates and the black x indicates non-simultaneous estimates. The dashed line represents the equality of the estimates and the gray shaded area the $63\%$ on-average error from \cite{Liodakis2015-II}.}
\label{plt:DVAR_vs_simnotsim}
\end{figure}

We see a wide range of results for the sources where there are estimates in two frequencies. This can be attributed to the fact that, at least for one of the estimates in each case, we used radio observations at a frequency other than the turnover frequency of the synchrotron spectrum. It is interesting to note that there is no systematic trend, i.e., a higher Doppler factor estimate at higher frequencies or vise versa. This result emphasizes the importance of treating each source separately and performing the radio observations at the turnover frequency.
{It is clear that one of the two estimates is more reliable for 3C345  ($\delta_{IC}$ = 7.4 obtained at 43 GHz), since it is consistent with $\langle \delta_{var}\rangle$ and the alternative estimate is unrealistically high (114.8). In addition, the turnover varies between $15\leq$ and $\leq86$ GHz (Angelakis et al. in prep.), which is consistent with our results suggesting that the 43 GHz $\delta_{IC}$ is an accurate estimate of the Doppler factor.  A similar case is for 3C454.3. Although we cannot constrain the turnover for 3C454.3, it is obvious that one estimate (at 43~GHz) is unrealistically high (119.9), {while the second (at 15~GHz) is consistent with $\langle \delta_{var}\rangle$. The unrealistically high value at 43 GHz could also be attributed to the relatively large decrease in angular size compared to the size of the core at 15 GHz. If this is indeed the size of the core at 43 GHz, it would suggest a significant decrease in the size of the jet between the two regions probed by these frequencies. This apparent decrease in size could be the result of small scale jet variation (e.g., a bent jet  or precession), although we are not able to excluded data-related artefacts. For this reason, our results for the 3C454.3 at 43~GHz should be treated with caution.} 
} 

Given the typical span of reported estimates for blazar Doppler factors in the literature ($\sim$ 0-45) it is possible to have an offset from the ``true'' Doppler factor of an order of magnitude and still be within familiar limits. This might as well be the case for PKS1730-130 for which one estimate is 2.6 and the other 38.5. In addition neither estimate is in agreement with $\langle \delta_{var}\rangle$ or any of the individual $\delta_{var}$ estimates that lie in between those estimates. It is possible that the turnover is between 15 and 43 GHz so that {one frequency overestimates while the other underestimates the Doppler factor.} Results from F-GAMMA find that the turnover is changing, while residing $<86$ GHz, which does not provide any constrains on its position, and hence we cannot draw any firm conclusions.

{
Using these estimates we can roughly estimate (for this sample) the impact of not using the turnover frequency for IC Doppler factors. Taking the ratio of the two estimates ($\delta_{IC,high}/\delta_{IC,low}$), we find that the difference between estimates can be
as high as a factor of $\sim 15$.

}

{
We can distinguish two cases: a) all estimates of the two methods, i.e., this work, average of H09, and L17, are in agreement; and b) the estimates derived in this work do not agree with the $\langle \delta_{var}\rangle$ estimate.

\begin{itemize}

{\item {\bf a) All estimates of the two methods (this work and $\langle \delta_{var} \rangle $ ) are in agreement}. There are {five sources in this category: PKS1510-089, PKS1633+382, 3C273, 3C345, and 3C454.3.} Given the different assumptions and approaches involved in the two estimates, it is highly unlikely that some unknown common bias would lead to the same result. {Especially for 3C273 and 3C345, even the original  $\delta_{IC}$ estimate from G93 is in agreement with the $\langle \delta_{var} \rangle $ estimate}. Thus we can conclude that since all methods converge to the same result for these five sources, the  Doppler factor of the jet in the radio region can be effectively constrained to the value reported in Table \ref{tab:doppler_est}. 
}

{\item {\bf  b) The estimates derived in this work do not agree with the $\langle \delta_{var}\rangle$ estimate.} The sources in this category are  PKS0420-014, PKS0528+135, PKS1156+295, PKS1730-130, 3C279, and CTA102. The source 3C279 has an unrealistically high value, which is most likely due to the turnover effect. {This is also confirmed by the fact that the G93 estimate is consistent within errors with $\langle \delta_{var} \rangle $ as well as with H09 and L17 estimates individually. As discussed above, the turnover frequency is prone to change over time. For this  reason, it is possible that at the time of observations (when the G93 estimate was derived) the turnover frequency was close to 5~GHz (which is the observing frequency used in G93), while the turnover frequency at the time of the observations used in this work was far from 15~GHz}.

 {For  PKS1156+295, the $\delta_{IC}$ and $\langle \delta_{var} \rangle$  estimates do not converge and $\delta_{IC}$ is not within 1$\sigma$ from any individual $\delta_{var}$ estimate (3$\sigma$ away from the H09 estimate). It is interesting that  although the H09 estimate is relatively high (28.5) the re-estimated value (6.0) is more consistent with the G93 estimate (9.6) and the L17 (12.8) is relatively close, however the error estimate given in L17 is too small to account for the difference. The fact that the two  $\delta_{IC}$ estimates are consistent with each other reduces the likelihood of any turnover systematic effects since G93 and this work use different frequencies, while contamination of the X-ray flux from the big blue bump is unlikely \citep{Abdo2010-II}. The error estimate in L17 is derived through an iterative process where the best-fit model is altered both in flare rise time and flare amplitude, and by setting a limit to the maximum standard deviation of the residuals of the fitted light curve, these authors have acquired a range of acceptable models. The existence of a sharp flare in the light curve of PKS1156+295 sets a relatively rigid limit to the range of flare rise times, which results in a small error estimate (0.3\% percentage error) that is not necessarily accurate because in this case other, usually subdominant and ignored sources of error, may become more important\footnote{See, e.g., \cite{Cyburt2001} and \cite{Mouschovias2010} for a discussion of error underestimation due to usually subdominant error sources}. If instead we use the 16\% on average (for that sample) quoted in L17, the error estimate for this source becomes $\sigma_{\delta_{var}}=2.0$ which would bring both $\delta_{IC}$ estimates within 3$\sigma$. It is then possible that the uncertainty of the $\delta_{var}$ in L17 is underestimated and the true value of the Doppler factor for this source lies between 6.0 and 12.8.  
}

{ Similarly, for CTA102, the} re-estimated $\delta_{IC}$ is not consistent with the $\langle \delta_{var} \rangle$ estimate and is instead consistent with the estimate from G93. Here there is also no prominent big blue bump \citep{Fromm2015} and we can rule out any problems with the individual $\delta_{var}$ estimates since they are in very good agreement with each other. This case would indicate either significant external Compton flux or an additional unidentified source of error beyond those discussed in this work.

}
\end{itemize}

There are estimates derived in this work  that agree with one of the $\delta_{var}$ estimates. These sources are PKS0420-014 and PKS0528+135. All the estimates in agreement with H09 are above the pileups seen in the distribution of Doppler factors in the H09 sample (see \citealp{Liodakis2015-II}) and should not, in principle, be affected by the cadence of observations, which is the dominant source of systematics in the $\delta_{var}$ method.

Possible interpretations of the discrepancies between the estimates in H09 and L17 (for example variability of $\delta$) are discussed in L17. Unless an unknown cause has lead to the false estimation of the $\delta_{IC}$ derived in this work, agreement with an estimate from one approach and not the other should, in principle, weigh in favor of that $\delta_{var}$ estimate being a better representation of the  Doppler factor.

}

Another potential source of error is the imperfect simultaneity of the radio and X-ray observations. The observations used in this work are taken to be less than a week apart. This limit was set empirically given the typical X-ray and radio variability of LSP sources. To test whether this assumption would cause any shift in our estimates, we used PKS0528+134 for which there exist multiple quasi-contemporaneous observations. Keeping the date of the radio observation constant and letting that of the X-rays vary, we re-estimated the $\delta_{IC}$ for four different time intervals.

{Figure \ref{plt:frac_dif} shows various $\delta_{IC}$ estimates versus time separation between observations for PKS0528+134. Even with observations obtained one week apart, the results are within the typical span of Doppler factor estimates (see H09 and L17). When, however, the separation of the observations is more than a month apart, the values for the Doppler factor become unrealistically high. Keeping in mind that this limit could be different for every source, it is possible that for some of our sources the radio and X-ray observations are not ``simultaneous enough''. A more variable source in X-rays might have a smaller time tolerance between observations. This might constitute a problem especially in cases where observations are performed during outbursts (a very common observational bias), in which case the variability in either bands could increase.}

We finally test whether lack of simultaneity between X-ray and radio observations is responsible for most of the statistical error in $\delta_{IC}$ once all other sources of { systematic} error have been accounted for. We used the sources for which the estimates derived in this work are in agreement with the $\langle\delta_{var}\rangle$ estimates. We then tampered with our estimates, violating the simultaneity criterion. We use the same radio data, i.e., flux density, frequency, and angular diameter, but instead of the X-ray flux densities from Table \ref{tab:flux_values}, we used the corresponding values quoted in G93.

{
Figure \ref{plt:DVAR_vs_simnotsim} shows the comparison between $\langle\delta_{var}\rangle$ and the simultaneous and non-simultaneous estimates using various X-ray flux densities. The gray band around the dashed line represents the 63\% on-average uncertainty associated with $\delta_{IC}$ from G93 {\citep{Liodakis2015-II}}.  It is clear that the simultaneous estimates (green $\star$) fare much better than the non-simultaneous estimates (black x).} We can also roughly estimate the contribution of the lack of simultaneity to the statistical error by examining the fractional difference between the $\delta_{IC}$ from this work with the $\delta_{IC}$ from G93. { We find that the percentage error is between $\sim$3\% to $\sim$540\% with an average of $\sim$100\% using all of the estimates; for
sources with multiple estimates we use the one closest to a $\langle \delta_{var}
\rangle$ estimate.  If we only use the  estimates that are consistent with either $\langle \delta_{var} \rangle$ or at least one of the individual $\delta_{var}$ estimates, we find the percentage error to be between $\sim$3\% and $\sim$97\% with an average of $\sim$68\%, { which is consistent with the on-average population estimate (63\%) from \cite{Liodakis2015-II}}. We conclude that lack of simultaneity between X-ray and radio observations is the dominant contributor to the overall statistical error budget.
}

\section{Summary}\label{summary}

Population models have shown that the inverse-Compton and variability Doppler factor methods can describe FSRQs as a population, even though they are inconsistent on a source-by-source basis. In this work, we attempted to resolve this discrepancy by identifying and examining potential sources of systematic and statistical error involved in the inverse-Compton Doppler factor method, and by eliminating them, we provide a much more accurate estimate of the ``true'' Doppler factor of blazar jets to serve as a gateway to their, yet to be explored, rest frame. 

To that end, we re-estimated the $\delta_{IC}$ for 11 sources and compared them with {the average of two} independent estimates of $\delta_{var}$ following various approaches. Our findings can be summarized as follows:
\begin{enumerate}

\item{Careful sample selection is crucial for the correct application of the method, as application of the method to sources where the X-ray flux is contaminated by non-SSC sources can cause systematic shifts in the estimates of $\delta_{IC}$. Such sources are high synchrotron-peaked blazars; sources with a significant external Compton X-ray flux; or sources with a prominent big blue bump, although this is not always prohibitive. An example is PKS1510-089, where there is a visible big blue bump \citep{DAmmando2009}, but it does not extend to high energies and thus has no contribution to the X-ray flux. Evidence of this is the fact that all the available estimates for the Doppler factor for PKS1510-089 converge to the same result.}

\item{Radio observations at the synchrotron turnover frequency are an important aspect of the IC method. Since the turnover frequency is likely to change over time or during events to ensure robust results, it is a condition that should not be violated. Departure from the turnover frequency could result in a Doppler factor error as high as a factor of $\sim 15$.}

\item{Errors due to the lack of simultaneity between X-ray and radio observations constitute the dominant fraction of the statistical error in IC Doppler factors. Although the exact amount that can be accounted for varies from source to source, this value is on average $\sim68$\%, which is consistent with the average estimate from \cite{Liodakis2015-II}. The time delay limit of seven days set in this work works reasonably well for most sources, although its applicability depends on the source and time of observations (e.g., observations during outbursts). Thus variability in both bands (X-ray and radio) should be taken into account when choosing the maximum reliable time separation between radio and X-ray observations.}

\item{By resolving the discrepancies between all methods, we were able to effectively constrain the Doppler factor for five sources {(about 45\% of our sample), namely: 3C273}, 3C345, 3C454.3, PKS1510-089, and PKS1633+382, where the re-estimated $\delta_{IC}$ and the $\langle \delta_{var} \rangle$ converge to the same result. This not only gives us confidence in our analysis, but it also provides a strong case supporting that this is indeed the value of the ``true'' Doppler factor of each jet.}

\end{enumerate}

{Throughout this work we have assumed that the Doppler factor does not vary significantly with time. This is not necessarily true. We have already discussed that the differences between the two $\delta_{var}$ estimates (H09, L17) can arise from the different time span of observations of the two datasets (see Section \ref{error_analysis}). This can also be true for the $\delta_{IC}$ estimates. Jet precession, bent jets, or local acceleration or deceleration of jet components \citep{Lister2009-2,Homan2009,Homan2015} might contribute to the overall uncertainty of estimating the ``true'' Doppler factor. For these reasons, the comparison of the various Doppler factor methods should be in principle performed using contemporaneous datasets although this is not always feasible given the availability of data. It is possible that such phenomena are responsible for the unidentified source of error for CTA102 as well as the remaining sources that did not converge.}

{We were able to constrain 5 out of the 11 sources in our sample using the only available archival data for which systematics effects, such as the choice of observing frequency with respect to the turnover, could not be fully accounted for. New contemporaneous radio and X-ray observations following the guidelines described in this work will in principle constrain the Doppler factor of blazar jets with higher efficiency for a larger number of sources or alternatively reveal new aspects of blazar emissions mechanisms and jet processes. }

\begin{acknowledgements}

{The authors would like to thank Bia Boccardi, Nicola Marchili, and the anonymous referee for comments and suggestions that helped improve this work.} This research was supported by the ``Aristeia'' Action of the  ``Operational Program Education and Lifelong Learning'' and is co-funded by the European Social Fund (ESF) and Greek National Resources, and by the European Commission Seventh Framework Program (FP7) through grants PCIG10-GA-2011-304001 ``JetPop'' and PIRSES-GA-2012-31578 ``EuroCal''. AZ acknowledges funding from the European Research Council under the European Union's Seventh Framework Programme (FP/2007-2013)/ERC Grant Agreement n. 617001. T. Hovatta was supported by the Academy of Finland project number 267324. This research has made use of data from the MOJAVE database that is maintained by the MOJAVE team \citep{Lister2009}. This study makes use of 43 GHz VLBA data from the VLBA-BU Blazar Monitoring Program (VLBA-BU-BLAZAR;
http://www.bu.edu/blazars/VLBAproject.html), funded by NASA through the Fermi Guest Investigator Program. The VLBA is an instrument of the National Radio Astronomy Observatory. The National Radio Astronomy Observatory is a facility of the National Science Foundation operated by Associated Universities, Inc. 
\end{acknowledgements}

\bibliographystyle{aa}
\bibliography{bibliography} 

\label{lastpage}

\end{document}